\RequirePackage{ifpdf}
\ifpdf 
\documentclass[pdftex]{sigma}
\else
\documentclass{sigma}
\fi

\begin{document}

\allowdisplaybreaks

\renewcommand{\thefootnote}{$\star$}

\renewcommand{\PaperNumber}{017}

\FirstPageHeading

\ShortArticleName{Comments on the Dynamics of the Pais--Uhlenbeck Oscillator}

\ArticleName{Comments on the Dynamics\\ of the Pais--Uhlenbeck Oscillator\footnote{This paper is a contribution to the Proceedings of the VIIth Workshop ``Quantum Physics with Non-Hermitian Operators''
     (June 29 -- July 11, 2008, Benasque, Spain). The full collection
is available at
\href{http://www.emis.de/journals/SIGMA/PHHQP2008.html}{http://www.emis.de/journals/SIGMA/PHHQP2008.html}}}

\renewcommand{\thefootnote}{\arabic{footnote}}
\setcounter{footnote}{0}

\Author{Andrei V. SMILGA\,\footnote{On leave of absence from ITEP, Moscow, Russia.}}

\AuthorNameForHeading{A.V. Smilga}

\Address{SUBATECH, Universit\'e de Nantes,  4 rue Alfred Kastler, BP 20722, Nantes  44307, France}
\Email{\href{mailto:smilga@subatech.in2p3.fr}{smilga@subatech.in2p3.fr}}

\ArticleDates{Received November 24, 2008, in f\/inal form February 05,
2009; Published online February 12, 2009}

\Abstract{We discuss the quantum dynamics of the PU oscillator, i.e.\ the system with
the Lagrangian
\begin{gather}
\label{PUosc}
L   =  \frac 12 \left[ \ddot {q}^2 - (\Omega_1^2 + \Omega_2^2) \dot{q}^2 + \Omega_1^2 \Omega_2^2 q \right]
\ \ ({\rm +\  nonlinear \ terms})  .
 \end{gather}
When $\Omega_1 \neq \Omega_2$, the free PU oscillator has a pure point  spectrum that is dense everywhere.
When $\Omega_1 = \Omega_2$, the spectrum is continuous, $E \in \{-\infty, \infty \}$.
The spectrum is not bounded from below,
but that is not disastrous as the Hamiltonian is Hermitian and the evolution operator is unitary.
Generically, the inclusion of interaction terms breaks unitarity, but in some special cases unitarity
is preserved.
We discuss also the nonstandard realization of the PU oscillator suggested by Bender and Mannheim, where
the spectrum of the free Hamiltonian is positive def\/inite, but wave functions
   grow exponentially for large real values of canonical coordinates. The free  nonstandard PU oscillator
is unitary at $\Omega_1 \neq \Omega_2$, but unitarity is broken in the equal frequencies limit. }


\Classification{70H50; 70H14}

\section{Motivation}
Mechanical and f\/ield-theory systems involving higher derivatives in the Lagrangian attracted
(or, better to say, reattracted) recently
some attention. In~\cite{TOE,benign} we put forward arguments that
  the Theory of Everything may represent a variant of
 supersymmetric higher-derivative theory living in f\/lat
higher-dimensional space. Our Universe is associated then with a 3-brane classical solution
in this theory (a kind of soap bubble embedded in the f\/lat higher-dimensional
bulk), while gravity has the status of ef\/fective theory in the brane world-volume.

The f\/irst paper where such theories were considered dates back to 1950~\cite{PU}. It was shown there
that higher-derivative theories are in many cases ghost-ridden, which may break unitarity and render
theory meaningless. Recently, it was understood, however, that there {\it are} higher-derivative systems
where the ghosts are ``benign''  such that unitarity is preserved.
An example of  nontrivial supersymmetric higher-derivative system of such
benign variety was constructed and studied in~\cite{Robert}.

In this note, we concentrate on the dynamics of the system (\ref{PUosc}) (it is the simplest example of
higher-derivative
mechanical systems introduced and studied f\/irst in the paper \cite{PU}, which is known by the name
{\it Pais--Uhlenbeck oscillator}). Recently, the interest to this problem has been revived
 \cite{benign,dno,Bolonek,DM,BM1,BM2}. In spite of that many salient features of the PU oscillator dynamics were revealed
in previous studies, a certain confusion persists now in the literature. This especially concerns the
{\it interpretation} of the results obtained.
 Thus, we found it useful to write a kind of
mini-review including all relevant previous results with inaccuracies corrected, as well as some new remarks.
There are basically three
new observations:
\begin{itemize}\itemsep=0pt
\item The statement of~\cite{dno} that the Hamiltonian of the oscillator in the equal frequency limit
represents a set of Jordan blocks of inf\/inite dimension is compatible with the statement that the spectrum
of such system is continuous if the limit $\Omega_1 \to \Omega_2$ is def\/ined in a natural way \cite{PU,Bolonek}.
We note that the wave functions of continuous spectrum can be represented as superpositions of non-stationary
solutions to the Schr\"odinger equation (non-stationary solutions are characteristic for
Jordan block Hamiltonians.)
 \item In contrast to what we suggested earlier \cite{dno}, the spectrum stays unbounded from below also when nonlinear
terms in equation~(\ref{PUosc}) are included. In most cases, the latter bring about the quantum
collapse phenomenon and  breaking of unitarity. Unitarity survives, however, when interaction has a certain special
form.
\item In contrast to what was written in~\cite{BM2}, the free equal-frequency PU
oscillator
in the nonstandard realization is not unitary in the  usual sense of this word.
 \end{itemize}

\section{Hamiltonian and its spectrum}\label{section2}

 The canonical Hamiltonian corresponding to the Lagrangian (\ref{PUosc}) can be obtained using Ostrogradsky's
method~\cite{Ostr}.
Due to the presence of extra derivatives, the phase space involves besides $(q, p_q)$ an extra canonical
coordinate
 \begin{gather*}
x= \dot q
 \end{gather*}
 and the corresponding canonical momentum $p_x$. The canonical Hamiltonian is
\begin{gather}
\label{Ham12}
H   =  p_q x + \frac {p_x^2}2 + \frac {\big(\Omega_1^2 + \Omega_2^2\big) x^2}2 - \frac {\Omega_1^2 \Omega_2^2 q^2}2  .
  \end{gather}
Indeed, if writing the canonical Hamilton equations of motion and excluding $p_q$, $x$, $p_x$, we arrive at the
equation
\begin{gather*}
q^{(4)} + \big(\Omega_1^2 + \Omega_2^2\big) \ddot q + \Omega_1^2 \Omega_2^2 q   =  0 ,
 \end{gather*}
which can on the other hand be directly derived from the Lagrangian (\ref{PUosc}).

Consider f\/irst the case of inequal frequencies and
assume for def\/initeness $\Omega_1 > \Omega_2$.
The Hamiltonian (\ref{Ham12}) can then be brought into diagonal form by the following canonical transformation
\begin{gather}
q  =  \frac 1\Omega_1 \frac {\Omega_1 X_2  -  P_1}{
\sqrt{\Omega_1^2 - \Omega_2^2}}  , \qquad
x = \frac {\Omega_1 X_1 - P_2}
{\sqrt{ \Omega_1^2 - \Omega_2^2}}
  ,\nonumber \\
p_x  =   \frac {\Omega_1 P_1 - \Omega_2^2 X_2}{
\sqrt{\Omega_1^2 - \Omega_2^2}}
 , \qquad
p_q =  \Omega_1  \frac {\Omega_1 P_2  - \Omega_2^2 X_1}{
\sqrt{\Omega_1^2 - \Omega_2^2}}  . \label{canon}
 \end{gather}
The inverse of it is
 \begin{gather}
X_1 = \frac 1\Omega_1  \frac {p_q + \Omega_1^2 x }{
\sqrt{\Omega_1^2 - \Omega_2^2}} , \qquad
X_2 =  \frac {p_x + \Omega_1^2 q }{
\sqrt{\Omega_1^2 - \Omega_2^2}}  , \nonumber \\
P_1   =    \Omega_1 \frac {p_x + \Omega_2^2 q}{
\sqrt{\Omega_1^2 - \Omega_2^2}}
  , \qquad
P_2  =   \frac {p_q + \Omega_2^2 x}{
\sqrt{\Omega_1^2 - \Omega_2^2}}   .\label{canoninv}
 \end{gather}
We obtain
  \begin{gather}
\label{Hdiag}
H  =  \frac {P_1^2 + \Omega_1^2 X_1^2}2   -
\frac {P_2^2 + \Omega_2^2 X_2^2}2  .
  \end{gather}
The spectrum of this Hamiltonian is
\begin{gather}
\label{spec12}
E_{nm} = \left(n+ \frac 12 \right) \Omega_1 - \left( m + \frac 12 \right) \Omega_2, \qquad n,m = 0,1,2,\ldots.
\end{gather}

The eigenfunctions of the original Hamiltonian (\ref{Ham12}) are \cite{dno}
\begin{gather}
 \label{Psi}
\Psi_{nm}(q,x)\ = \ C_{nm} e^{-i\Omega_1 \Omega_2 qx} \exp\left\{ - \frac \Delta 2 \left( x^2 +
\Omega_1 \Omega_2 q^2 \right) \right\} \phi_{nm}(q,x) ,
 \end{gather}
where $\Delta = \Omega_1 - \Omega_2$ and
 \begin{gather}
\phi_{nm}(q, x)  =  \sum_{k=0}^m \left( \frac {i\Delta }{4 \sqrt{\Omega_1 \Omega_2}} \right)^k
\frac {m! (n-m)!}{(m-k)! k! (n-m+k)!} H^+_{n-m+k} H^-_k,\qquad  m\leq n   , \nonumber \\
\phi_{nm}(q, x)  =  \sum_{k=0}^n \left( \frac {i\Delta }{4 \sqrt{\Omega_1 \Omega_2}} \right)^k
\frac {n! (m-n)! }{(n-k)! k! (m-n+k)!} H^+_{k} H^-_{m-n+k},\qquad  m >  n .  \label{sumHerm}
 \end{gather}
Here $H^\pm_l$ are the Hermite polynomials of the following arguments,
 \begin{gather*}
H_l^+   \equiv \ H_l[i\sqrt{\Omega_1} (\Omega_2 q - ix)], \qquad
H_l^-   \equiv \ H_l[\sqrt{\Omega_2} (\Omega_1 q  + ix)]
 \end{gather*}
[$H_0(z) \equiv 1, H_1(z) \equiv  2z, H_2(z) \equiv  4z^2-2, \ldots$],
 and $C_{mn}$ are the irrelevant for our purposes normalization factors.

The wave functions (\ref{Psi}) are normalized implying that the spectrum is pure point\footnote{Mathematicians say that a self-adjoint operator $H$ has pure point spectrum  if all eigenfunctions of $H$ belong to ${\cal L}_2$, forming a
complete orthogonal basis there.}.
 However,
for generic incommensurable frequencies, the spectrum is  dense everywhere: however small  the interval
$\{E, E+dE\}$ is, it contains  an inf\/inite number of eigenvalues, and this is true for any~$E$. Obviously, this unusual property
is related to the fact that the spectrum has no bottom. We want to emphasize, however, that, though
this quantum system is unusual, it is not sick: the Hamiltonian~(\ref{Hdiag}) is Hermitian and the corresponding
evolution operator is unitary.

In the limit $\Omega_1 \to \Omega_2$, the transformation (\ref{canon}) becomes singular. To understand what
happens in this limit, it is best to look directly at the spectrum~(\ref{spec12})  and the wave functions
(\ref{Psi}), (\ref{sumHerm}).

If setting formally $\Omega_1 = \Omega_2$ in equation~(\ref{spec12})\footnote{See, however, the discussion below.}, we obtain
 \begin{gather}
\label{Enmdegen}
E_{nm} = \Omega(n-m)
 \end{gather}
meaning the inf\/inite degeneracy of the spectrum at each level. The wave functions are reduced to
 \begin{gather}
\label{Psidegen}
 \Psi_{nm}(q,x)\ \propto \ \left[
\begin{array}{c}  e^{-i\Omega^2 qx} H^+_{n-m},\qquad m \leq n,\vspace{1mm} \\
e^{-i\Omega^2 qx} H^-_{m-n},\qquad m > n.  \end{array} \right.
 \end{gather}
We see that, at each level with a given $n-m$, an inf\/inite dimensional {\it Jordan block}
appears \cite{dno}.

\section{Jordan blocks and continuous spectrum. Unitarity}\label{section3}

The presence of Jordan blocks in the Hamiltonian\footnote{The points in the parameter space where such Jordan blocks emerge are called
{\it exceptional} points, see e.g.~\cite{Hess} and references therein.}
implies usually the loss of unitarity. Indeed, consider the matrix Hamiltonian
  \begin{gather}
\label{HJordan}
H = \left( \begin{array}{cc}1&1 \\ 0&1  \end{array} \right)
 \end{gather}
 (it has only one eigenvector
$\psi = \left( \begin{array}{c}1 \\ 0 \end{array} \right) $ with eigenvalue 1). It is straightforward to see that the
time-dependent Schr\"odinger equation
 \begin{gather}
\label{Schr}
i \frac {d\psi}{dt}   = H\psi
\end{gather}
has the following general solution,
\begin{gather}
\label{solab}
\psi(t)   =  a \left( \begin{array}{c}1 \\ 0 \end{array} \right)e^{-it} +
b \left( \begin{array}{c} -it \\ 1 \end{array} \right)e^{-it}   .
 \end{gather}
When $b \neq 0$, the norm of $\psi(t)$ grows with time. The latter statement
 applies to the natural norm $\|\psi\| = \psi^\dagger \psi$
and also to any other positive def\/inite norm $ \sim \psi^\dagger M \psi$. If the norm is not positive
def\/inite, it is not true, as is clearly seen when choosing
\[
M =   \left( \begin{array}{cc}0&0 \\ 0&1  \end{array} \right)  .
\]
Such a norm is degenerate, however. It projects the full 2-dimensional Hilbert space where the Hamiltonian
(\ref{HJordan}) is def\/ined to
  a one-dimensional subspace $\psi =  \left( \begin{array}{c}0 \\ c \end{array} \right)$. The dynamics in this
subspace is unitary, the dynamics in  full Hilbert space is not.

This reasoning applies to any Jordan block of f\/inite dimension and to any Hamiltonian inclu\-ding such.
However, in our case, the dimensions of the Jordan blocks are {\it infinite}. A~remarkable fact is
that in this case unitarity is not necessarily broken. In particular, a unitary evolution operator
can well be def\/ined for the free PU oscillator at equal frequencies. The novel feature compared
to the inequal frequencies case is the appearance of continuous spectrum \cite{PU,Bolonek}.

Strong indications that the spectrum is continuous follow already from inspecting the
spectrum (\ref{spec12}),  the wave functions (\ref{Psi}), and their fate in the limit $\Omega_1 \to
\Omega_2$.
  \begin{itemize}\itemsep=0pt
\item One can notice f\/irst of all that the discrete spectrum (\ref{Enmdegen}) is only obtained from
(\ref{spec12}) if setting $\Omega_1 = \Omega_2$ while keeping $n$, $m$ f\/inite. It {\it is} possible to take
the limit in such a way (this implies a kind of ultraviolet regularization), but a more natural approach
is to allow~$n$,~$m$ to be arbitrary large. When $\Delta = \Omega_1 - \Omega_2$ is sent to zero and $n,m$ are sent
to inf\/inity such that $n-m$ is kept f\/ixed and $n\Delta$ is kept f\/inite, the energy (\ref{spec12})
can acquire an arbitrary value and not only the discrete values (\ref{Enmdegen}).
 \item
In the limit $\Delta \to 0$, the wave functions (\ref{Psi}) are no longer normalizable, and this
suggests that the spectrum is continuous.
\end{itemize}

Let us now prove the continuity of the spectrum by constructing explicitly the wave functions
of arbitrary energy $E$.
At the f\/irst step, let us get rid of the terms $\propto x^2$ and $\propto q^2$ in the Hamiltonian by representing
an eigenfunction of (\ref{Ham12}) as
$\Psi(q,x) = e^{-i\Omega^2 xq} \phi(q,x)$. The operator acting on $\phi(q,x)$ is
  \begin{gather}
\label{Htild}
\tilde{H}   =  \frac 12 p_x^2 + xp_q - \Omega^2 q p_x   .
 \end{gather}
It is convenient then to perform the canonical transformation \cite{PU,Bolonek}
 \begin{gather}
\label{canonH}
p_x \to p_x, \qquad p_q \to \Omega p_q, \qquad x \to x+ \frac 1{4\Omega} p_q, \qquad
q \to \frac 1\Omega q +  \frac 1{4\Omega^2} p_x   .
 \end{gather}
giving the new Hamiltonian
 \begin{gather}
\label{Hprime}
H'  =  \frac {p_x^2+p_q^2}4 + \Omega (xp_q - qp_x)  .
  \end{gather}
The transformation (\ref{canonH}) is the superposition of the scale transformation\footnote{It amounts to a unitary transformation with
$S = e^{-i(qp_q + p_q q) \ln \Omega/2}$, but  this representation is not particularly
useful.}
 $q \to q/\Omega$, $p_q \to \Omega p_q$
and the unitary transformation $O \to e^ROe^{-R}$ with $R = ip_x p_q/(4\Omega)$.
The eigenfunctions of the Hamiltonian
(\ref{Htild}) is obtained from  the eigenfunctions of the Hamiltonian (\ref{Hprime}) as\footnote{For a general theory of quantum canonical transformations see \cite{Anderson} and references therein.}
\[
\phi(q,x)   =  \exp\left\{ \frac i{4\Omega^2} \frac {\partial^2}{\partial x \partial q}
 \right\} \phi'(\Omega q, x)  .
 \]
The eigenfunctions of $H'$ are known,
 \begin{gather*}
\phi'_{lk}(q', x; t) \propto J_l(kr)e^{-il\theta}\, e^{-it(l\Omega + k^2/4)}   ,
 \end{gather*}
where $(r,\theta)$ are the polar coordinates in the plane $(q',x)$ and
$l = 0, \pm 1, \pm 2, \ldots $ are the eigenvalues
of the angular momentum operator ${L} = xp'_q- q'p_x$.
We obtain
 \begin{gather*}
\phi_{lk} (q,x; t)   \ \propto \ \exp \left\{  \frac i{4\Omega^2} \frac {\partial^2}{\partial x \partial q} \right\}
\left[ J_l\left( k \sqrt{x^2 + \Omega^2 q^2} \right) \left( \frac {\Omega q - ix}{\Omega q + ix} \right)^{l/2} \right]
e^{-it(l\Omega + k^2/4)}   .
 \end{gather*}
Introducing
\begin{gather*} z = \sqrt{\Omega} (x+i\Omega q) , \qquad
  w = i\bar z =  \sqrt{\Omega} (ix+\Omega q)  ,
\end{gather*}
expanding the Bessel function and using the identity
\[
 \exp\left\{ - \frac 14 \frac {\partial^2}{\partial z^2} \right\}   z^n
  =  2^{-n} H_n(z)   ,
\]
we f\/inally derive
 \begin{gather}
\label{Psilk}
\Psi_{lk} \ \propto \ e^{-it(l\Omega + k^2/4)}
e^{-i\Omega^2 qx} \sum_{m=0}^\infty \left( \frac {i k^2}\Omega \right)^m
\frac {H_{l+m}(z) H_m(w)}{4^{2m+l} m!(l+m)!}  .
 \end{gather}
The structure of this expression is similar to that in equations~(\ref{Psi}), (\ref{sumHerm}), but the meaning is dif\/ferent:
the expansion goes now in the spectral parameter $k^2/\Omega$ rather than in
the parameter of the Hamiltonian $\Delta/\sqrt{\Omega_1 \Omega_2}$ as in equation~(\ref{sumHerm}). In addition, equation~(\ref{Psilk})
represents an inf\/inite series rather than a f\/inite sum.
 We can note that each level in the spectrum  is inf\/initely degenerate. The eigenfuctions $\Psi_{lk}$,
 $\Psi_{l-1,\sqrt{k^2 + 4\Omega}}$, etc. have the same energy.
This is {\it not} a Jordan degeneracy discussed above: the basis (\ref{Psilk}) diagonalizes the Hamiltonian,
 and the functions $\Psi_{lk}$ with dif\/ferent $l$ are distinguished by the eigenvalue of the ``angular
momentum'' operator,
 \begin{gather*}
L = \frac {xp_q}{2\Omega} - \frac \Omega 2 qp_x + \frac 1{4\Omega} \left( p_x^2 - \frac {p_q^2}{\Omega^2} \right)
+ \frac {3\Omega x^2}4 - \frac {3\Omega^3 q^2}4
\end{gather*}
(we have rotated the operator  ${L} = xp'_q - q'p_x$ back to original variables), that commutes with the Hamiltonian~(\ref{Ham12}).

We have made two seemingly conf\/licting statements: $(i)$ the spectrum of the Hamiltonian involves a set
of inf\/inite-dimensional Jordan blocks and $(ii)$ the spectrum is continuous. How to reconcile them?
To understand it better, consider the trivial free Hamiltonian
\[
H   = - \frac 12 \frac {\partial^2}{\partial x^2}
\]
with the eigenfunctions
 \begin{gather}
\label{Psifree}
 \Psi(x; t)   =  \exp\left\{ ikx - \frac {ik^2}2 t \right\}  .
 \end{gather}
Note now that not only (\ref{Psifree}), but also every term of its expansion in $k$,
\begin{gather}
k^0  :\ \ \Psi_0(x;t)  =  1   , \nonumber \\
k^1  :\ \ \Psi_1(x;t)  =  x   , \nonumber \\
k^2  :\ \ \Psi_2(x;t)  =  t - ix^2   , \nonumber \\
k^3  :\ \ \Psi_3(x;t)  =  xt - \frac {ix^3}3   , \nonumber \\
k^4  :\ \ \Psi_4(x;t)  =  t^2 - 2itx^2 - \frac {x^4}3    ,\label{kuski}
 \end{gather}
etc.,  satisfy the time-dependent Schr\"odinger equation (\ref{Schr}). The functions (\ref{kuski}) represent polynomials
 in $x$ and, what is
especially noteworthy, in $t$, much similar to non-stationary solutions~(\ref{solab}) characteristic to Jordan block Hamiltonians.

On the other hand, the functions~(\ref{kuski}) grow at large $x$, are not normalizable, and do not form a~basis of a reasonable
Hilbert space. The only way to def\/ine the latter is to sum over all $\Psi_n(x;t)$ with a proper weight and go back to the standard
continuous spectrum wave functions~(\ref{Psifree}), which can be dealt with by introducing a  box of f\/inite length~$L$
(where the def\/inition of Hilbert space presents no problem) and sending then~$L$ to inf\/inity.

By the same token, individual terms of the expansion of (\ref{Psilk}) in $k$ represent nonstationary solutions
of the Schr\"odinger equation with the Hamiltonian (\ref{Ham12}). For $l=0$, these solutions are
\begin{gather}
 k^0 : \ \ e^{-i\Omega^2 qx}  , \nonumber \\
 k^2 : \ \ \big[t - i\big(x^2 + \Omega^2 q^2\big)\big]  e^{-i\Omega^2 qx} , \nonumber \\
 k^4 : \ \ \left[ t^2 - 2it(x^2 + \Omega^2 q^2) - \frac {(\Omega^2 q^2 + x^2)^2}2 -
iqx + \frac 1{8\Omega^2} \right]  e^{-i\Omega^2 qx} ,\label{kuskiPU}
 \end{gather}
etc. The f\/irst function in (\ref{kuskiPU}) coincides with the wave function (\ref{Psidegen}) with $n-m = 0$. The higher
functions may be interpreted as its non-stationary ``descendants'' associated with the presence of the inf\/inite-dimensional
Jordan block at the level $E_{nm} = 0$.

However, as was also the case in the previous example, these descendants grow at large~$x$,~$q$ and only their superpositions
(\ref{Psilk}) form a proper Hilbert space basis.

\section{Including interactions}\label{section4}

We have learned in the previous section that, in spite of unusual features (the absence of the ground state), the free PU
oscillator represents a benign unitary quantum system both when $\Omega_1 \neq \Omega_2$ and when $\Omega_1 = \Omega_2$.
It is especially clear in the inequal frequencies case when the Hamiltonian can be brought in the form (\ref{Hdiag}).
When two dif\/ferent oscillators are decoupled, it does not matter much that the energies of one of the  oscillators
are counted with the negative sign. Each oscillator lives its own life and the energy sign is basically a bookkeeping
issue.

The danger arises when the oscillators start to interact. The subsystem 1 can give energy to subsystem 2 such that the
absolute values of the energies of both subsystems increase. This brings about a potential instability
that may lead to collapse and loss of unitarity. We will see that, generically, such collapse occurs, indeed. But not always.
When the interaction Hamiltonian has some special form, there is no collapse and unitarity is preserved.

Consider the Hamiltonian (acting on $\phi(q,x)$)
  \begin{gather}
\label{HPUint}
H   =  - \frac 12 \frac {\partial^2}{\partial x^2} + i \Omega^2 q \frac \partial {\partial x} -
ix \frac \partial {\partial q} + \alpha q^4 + \beta q^2 x^2 + \gamma x^4  .
 \end{gather}
Let us f\/ind out f\/irst whether the spectrum is still unbounded from below as
 for the free PU oscillator. The answer is
--- yes, it is. To see that, choose the variational Ansatz
 \begin{gather}
\label{Ansatz}
\phi(q,x)   =  \exp\left\{ - \frac {Aq^2}2 - iBxq -  \frac {Cx^2}2 \right\} \sqrt{ \frac {AC} \pi }  .
 \end{gather}
The variational energy is
 \begin{gather*}
E(A,B,C)   =  \frac C4 + \frac {B^2}{4A} + \frac {B\Omega^2}{2A} - \frac B{2C} + \frac {3\alpha}{4A^2}
+ \frac {3\gamma}{4C^2} + \frac \beta {4AC}   .
  \end{gather*}
 By varying $A$, $B$, $C$, one can make it arbitrarily negative. Indeed, choose f\/irst $A$ very large such that
\[
 E( \infty,B,C) =   \frac C4  - \frac B{2C} + \frac {3\gamma}{4C^2} .
 \]
Keeping $C$ f\/inite and increasing $B$, we can make $-E$ arbitrary large.
In~\cite{dno}, we used a dif\/ferent variational Ansatz with which the energy of the ground state
could not be made arbitrarily negative. However, as  the energy can be lowered
indef\/initely with a more general Ansatz~(\ref{Ansatz}), we conclude that the ground state is absent.

When $\alpha$ or $\beta$ or $\gamma$ are nonzero, the classical dynamics of the Hamiltonian (\ref{HPUint}) involves collapse.
In the certain region of parameters, there is an island of stability around the perturbative vacuum \cite{benign},
def\/ined as  the point $q=\dot{q} = \ddot{q} = q^{(3)} = 0$. With initial conditions at the vicinity of this point,
 the trajectories display benign behaviour peacefully
oscillating near the origin. Generically, however, the trajectories go astray reaching inf\/inity in f\/inite time.

The simplest well-known system displaying such behaviour describes 3D motion of the particle
 with the attractive potential
$V(r) = -\kappa/r^2$. Classically, for certain initial conditions, the particle  falls to the center in f\/inite time.
 The quantum dynamics
of the system depends in this case on the value of $\kappa$. If $m\kappa < 1/8$, the ground state exists
and unitarity is preserved. On the other hand, if $m\kappa > 1/8$, the spectrum is not bounded from below and, which is worse,
the spectral problem cannot be well posed until some extra boundary conditions are set at the vicinity of the origin.
The spectrum depends then on these boundary conditions \cite{fall}.
Setting such boundary conditions is tantamount to regularizing
the singularity of the potential (the spectrum thus depends on the way it is done).
Without such regularization, the probability  leaks out through the singularity and unitarity
is violated.
One can conjecture that the presence of classical collapsing trajectories {\it and} the absence of the ground
state always imply violation of unitarity. We do not know whether a mathematical theorem to this ef\/fect can be proven, but,
heuristically, it is suggestive. Indeed, if the spectrum involves the states with arbitrary low energies, the corresponding wave
functions should have main support near the singularity, where theory needs regularization to prevent the leakage of
probability.

\begin{figure}[t]
\centerline{\includegraphics[width=4.0in]{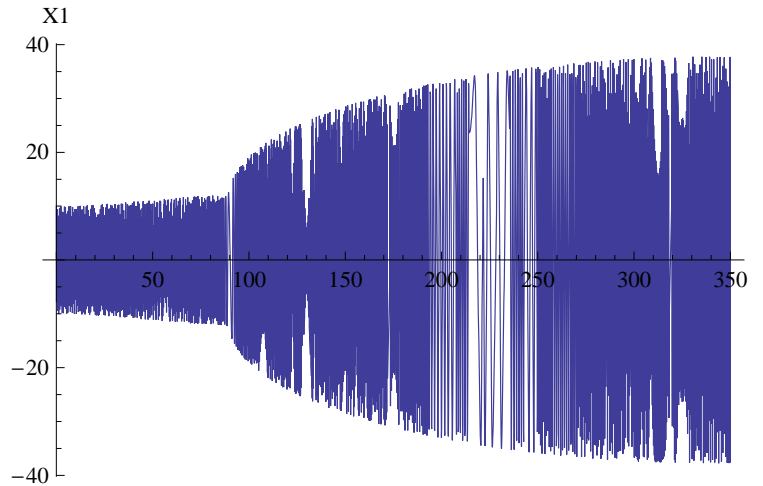}}
\caption{A typical classical trajectory of the Hamiltonian~(\ref{H12nonlin}).}
\label{oblastx3}
\end{figure}

In other words, the nonlinear PU oscillator is in most cases sick: collapsing classical trajectories lead to
the quantum collapse. However,  in some cases there is no collapse.
Consider the PU oscillator with inequal frequencies and let us add to the Hamiltonian (\ref{Hdiag}) the potential
 \begin{gather}
\label{VX12}
V(X_1, X_2)   =  \lambda (X_1 - X_2) (X_1 + X_2)^3, \qquad \lambda > 0.
 \end{gather}
The numerical solution to the classical equations of motion,
 \begin{gather*}
\ddot{X}_1 + \Omega_1^2 X_1 + \frac {\partial V}{ \partial X_1} = 0 \nonumber  , \\
\ddot{X}_2 + \Omega_2^2 X_2 - \frac {\partial V}{ \partial X_2} = 0  ,
 \end{gather*}
do not display any collapse. The amplitude of oscillations somewhat grows with time,
 but this growth is rather smooth
(see Fig.~\ref{oblastx3}).

The Hamiltonian
\begin{gather}
\label{H12nonlin}
H   =  \frac {P_1^2 + \Omega_1^2 X_1^2}2   -
\frac {P_2^2 + \Omega_2^2 X_2^2}2 + \lambda (X_1-X_2) (X_1+ X_2)^3
 \end{gather}
is a close relative to the supersymmetric Hamiltonian considered in~\cite{Robert}.
The bosonic part of the latter is
 \begin{gather}
\label{HRobert}
H   =  pP + D(\Omega^2 x + \lambda x^3) ,
 \end{gather}
where $(p,P)$ are the canonical momenta of $(x,D)$. By introducing $X_{1,2} = (x \pm D)/\sqrt{2}$,
it is reduced
to the Hamiltonian (\ref{H12nonlin}) with $\Omega_1 = \Omega_2 = \Omega$.\footnote{To avoid confusion, it is worth  reminding that the limit $\Omega_1 \to \Omega_2$ is singular
and the Hamiltonian (\ref{Hdiag}) with $\Omega_1 = \Omega_2$ does {\it not} describe the PU oscillator
with equal frequencies.} The system (\ref{HRobert}) is integrable and the
 classical trajectories are expressed analytically into Jacobi elliptic
functions. In this case, the amplitude of oscillations grows linearly with time, but this does not represent
a disaster as the trajectories do not run into inf\/inity in a f\/inite time.
The Schr\"odinger equation for the Hamiltonian (\ref{HRobert}) can be solved analytically~\cite{Robert}.
The spectrum is continuous, with eigenvalues lying in two intervals $]-\infty, -\Omega]\cup [\Omega, \infty[$
plus the eigenvalue $\{0\}$. The quantum spectrum for the Hamiltonian (\ref{H12nonlin}) should have the same
qualitative features. Besides (\ref{VX12}), there are other interaction potentials providing for the same behaviour.
One of them is $V(X_1, X_2)  = \lambda (X_1 - X_2)^3 (X_1 + X_2) $.
 Note that,
being expressed in terms of the original variables $x$, $q$, $p_x$,~$p_q$, the Hamiltonian~(\ref{H12nonlin})
acquires a rather complicated form and the corresponding Lagrangian depending on $q$ and its time derivatives
is even more complicated. Natural nonlinear extensions of the Lagrangian (\ref{PUosc}) seem all to involve classical
and quantum collapse.

\section[Complexification]{Complexif\/ication}\label{section5}

The spectral Sturm--Liouville problem is def\/ined when the operator in question (Hamiltonian)
is given {\it and} boundary conditions are specif\/ied. It is possible thus to have dif\/ferent spectral problems
associated with a given Hamiltonian.

\begin{figure}[t]
\centerline{\includegraphics[width=2.0in]{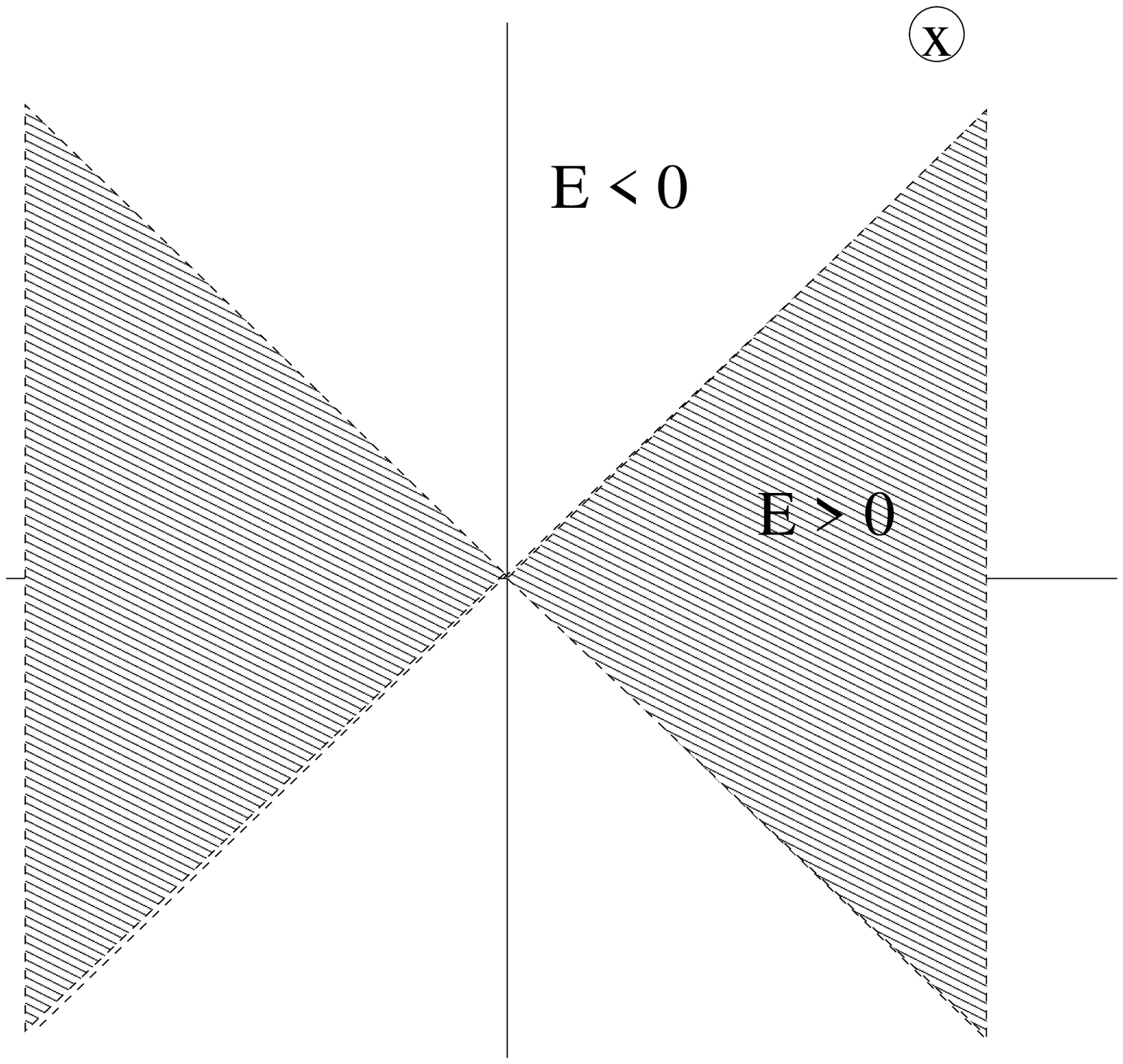}}
\caption{Stokes wedges.}\label{fig2}
\end{figure}

Take a simple oscillator,
\[
H  = \frac 12 (P^2 + \Omega^2 X^2) .
\]
If considering the spectral problem at real $X$ and requiring that wave functions vanish at
inf\/inity, we obtain the usual spectrum
\[
E_n = \Omega(n+ 1/2) .
\]
The wave functions,  $\Psi \propto e^{-X^2/2}$, continued
analytically to complex $X$,
 vanish asymptotically not only on the real axis, but in the whole {\it Stokes wedge} with $ |{\rm arg} (X)|  < \pi/4$.
Nothing prevents, however, to consider another
spectral problem and require that wave functions vanish at large {\it imaginary} $X$.
The corresponding eigenfunctions are $\Psi_n \propto H_n(iX) e^{X^2/2}$. They vanish in the complementary Stokes wedge
as shown in Fig.~\ref{fig2}. The spectrum is now
\[
E_n = -\Omega(n+ 1/2) .
\]

We have seen that the free PU oscillator with dif\/ferent frequencies is equivalent to the combination of two oscillators
(\ref{Hdiag}) with the spectrum (\ref{spec12}). But the latter is true only if $q$, $x$, $p_q$, $p_x$ and hence $X_{1,2}$ are assumed to be real.
One can on the other hand assume that, while $X_1$ is kept real, $X_2$ is imaginary. In this case, the sign of the energy eigenvalues
of the second oscillator is reversed, and one obtains a positive def\/inite spectrum
 \begin{gather}
\label{specrot12}
E_{nm} = \left(n+ \frac 12 \right) \Omega_1 + \left( m + \frac 12 \right) \Omega_2,\qquad  n,m = 0,1,\ldots
\end{gather}
with no trace of ghosts. Another way to obtain this result is the following  \cite{BM1,BM2}. Let us consider
the Hamiltonian (\ref{Ham12}) and assume $x$ being real and $q$ being imaginary there\footnote{From the viewpoint of the original Lagrangian (\ref{PUosc}), it is not so natural
to assume that the coordinate $q$ is imaginary, while its time derivative $x = \dot q$ is real.
But, formally, one is allowed to do it.}, $q = iy$, $p_q = -ip_y$. The Hamiltonian acquires the form
  \begin{gather}
\label{Hamrot}
H  =  \frac {p_x^2}2 - ixp_y  + \frac {(\Omega_1^2 + \Omega_2^2) x^2}2 + \frac {\Omega_1^2 \Omega_2^2 y^2}2 .
  \end{gather}
The second term in this expression is complex which may lead to worries whether the Hamiltonian (\ref{Hamrot}) makes sense.
This worries are not grounded, however. Though the Hamiltonian (\ref{Hamrot}) is not manifestly
Hermitian,  it belongs to the class of {\it crypto-Hermitian} Hamiltonians with {\it real} spectrum \cite{crypto}.
The Hamiltonian (\ref{Hamrot}) can be diagonalized
to the manifestly Hermitian form,
\begin{gather}
\label{Hdiagrot}
H_{\rm diag}   = \frac {P_1^2 + \Omega_1^2 X_1^2}2   +
\frac {P_2^2 + \Omega_2^2 X_2^2}2 .
  \end{gather}
by a complex canonical transformation (a brother of the transformation (\ref{canoninv})),
 \begin{gather}
X_1 \equiv \tilde{x}  =  \frac 1\Omega_1  \frac { \Omega_1^2 x - ip_y }{
\sqrt{\Omega_1^2 - \Omega_2^2}}  , \qquad
X_2 \equiv \tilde{y} = \frac { \Omega_1^2 y - ip_x }{
\sqrt{\Omega_1^2 - \Omega_2^2}}  , \nonumber \\
P_1 \equiv \tilde{p}_x   =    \Omega_1 \frac {p_x + i\Omega_2^2 y}{
\sqrt{\Omega_1^2 - \Omega_2^2}}
  , \qquad
P_2  \equiv \tilde{p}_y =   \frac {p_y + i\Omega_2^2 x}{
\sqrt{\Omega_1^2 - \Omega_2^2}}  .\label{canoncompl}
 \end{gather}
This transformation amounts to a certain similarity nonunitary transformation,
 \begin{gather}
\label{simil}
H_{\rm diag}    =   e^{-S} e^{-R} H e^{R} e^{S}
 \end{gather}
with
 \begin{gather*}
R = \left( \frac {p_x p_y}{2\Omega_1 \Omega_2}  + \frac {xy \Omega_1 \Omega_2}2
\right) \ln \frac {\Omega_1 + \Omega_2}{\Omega_1 - \Omega_2} ,
 \end{gather*}
and
\begin{gather*}
S =    \frac {i \ln \Omega_1 }2  (yp_y + p_y y)   .
  \end{gather*}

To f\/ind the eigenfunctions of (\ref{Hamrot}), one can either rotate away the eigenfunctions of (\ref{Hdiagrot}),
$\Psi = e^{R} e^{S} \Psi_{\rm diag}$ or proceed in the same way as in~\cite{dno}   representing  the wave functions
 as
\begin{gather*}
\Psi_{nm}   =  \phi_{nm} \exp \left\{ - \frac {\Omega_1 + \Omega_2}2 (x^2 + \Omega_1 \Omega_2 y^2) -
\Omega_1 \Omega_2 xy \right\}  .
 \end{gather*}
The operator acting on $\phi_{nm}(x,y)$ is
 \begin{gather*}
\hat O   =  -\frac 12 \frac {\partial^2}{\partial x^2} - x\frac \partial{\partial y} + [x(\Omega_1 + \Omega_2) + y\Omega_1 \Omega_2]
\frac \partial{\partial x} + \frac {\Omega_1 + \Omega_2}2  .
 \end{gather*}
Introducing
\begin{gather*}
z = \sqrt{\Omega_1} (x + \Omega_2 y), \qquad  w = \sqrt{\Omega_2}(x + \Omega_1 y)   ,
 \end{gather*}
it acquires the form
 \begin{gather}
\label{Ozw}
 \hat O   =  \Omega_1 \left[ -\frac 12 \frac {\partial^2}{\partial z^2} + z \frac \partial {\partial z} \right] +
\Omega_2 \left[ -\frac 12 \frac {\partial^2}{\partial w^2} + w \frac \partial {\partial w} \right] - \sqrt{\Omega_1 \Omega_2}
 \frac {\partial^2}{\partial z \partial w } + \frac {\Omega_1 + \Omega_2}2  .
 \end{gather}
The eigenfunctions of $\hat O$ are polynomials. When the dependence on the variable $w$ or on the variable $z$ is suppressed,
they are conventional Hermite polynomials,
\begin{gather*}
\phi_{n0}   =  H_n(z),\qquad \phi_{0m} = H_m(w) .
 \end{gather*}
When both $n \neq 0$ and $m \neq 0$, the eigenfunction is
 \begin{gather}
\label{sumHerm1}
\phi_{nm}(x,y) = \sum_{k=0}^{\min(n,m)} \left(- \frac {\Omega_1 + \Omega_2 }{4 \sqrt{\Omega_1 \Omega_2}} \right)^k
\frac {m! (n-m)!}{(m-k)! k! (n-m+k)!} H_{n-m+k}(z) H_k(w)  ,
 \end{gather}
The expression (\ref{sumHerm1}) is very similar to  (\ref{sumHerm}) and is obtained from the latter by substituting
$i\Delta = i(\Omega_1 - \Omega_2) \to -(\Omega_1 + \Omega_2)$ and  replacing the arguments $i\sqrt{\Omega_1} (\Omega_2 q - ix) \to z$, $\sqrt{\Omega_2} (\Omega_1 q + ix) \to w$.

There is a price, however, that one has to pay for getting rid of negative energies in the spectrum. The Hamiltonian (\ref{Hamrot})
{\it is} not Hermitian and this means that the conventionally def\/ined norm $\|\Psi\| = \iint  dx dy
\left|\Psi(x,y) \right|^2$
is not preserved during evolution. In addition, the eigenfunctions~(\ref{sumHerm1}) do not represent an
orthogonal basis with respect
to this norm. True, crypto-Hermiticity of the Hamiltonian implies that a norm with respect to which the evolution is unitary
can be def\/ined, but this norm,
\[
\| \Psi \|'    =  \langle \Psi^*  e^{-2R}  \Psi \rangle   ,
\]
has a complicated nonlocal structure.

The transformations (\ref{canoncompl}) are singular at $\Omega_1 = \Omega_2$. This suggests that,
as it was the case for the standard PU oscillator, the point $\Omega_1=\Omega_2$ is exceptional
also in the nonstandard realization. The emergence of Jordan blocks  in this limit was noticed
 back in~\cite{DM}, but  the best way to see that is to follow, as Bender and Mannheim did~\cite{BM2},
  the approach of~\cite{dno} and to explore the fate
of eigenfunctions (\ref{sumHerm1}) in the limit $\Omega_1 \to \Omega_2$.

In this limit, the variables $z$ and $w$ coincide. The eigenfunctions (\ref{sumHerm1}) for
a given $n+m$ also all coincide in this limit, $\phi_{nm}(z) \sim H_{n+m}(z)$. To derive the latter,
note that the operator (\ref{Ozw}) acquires in the limit $\Omega_1 \to \Omega_2$ the form
 \begin{gather*}
\hat O_{\Omega_1 = \Omega_2}   =   \Omega
\left[ -\frac {\partial^2}{\partial z^2} + 2z \frac \partial {\partial z} + 1
\right]
 \end{gather*}
and its eigenfunctions are simple Hermite polynomials, indeed. After massaging equation~(\ref{sumHerm1})
a little bit, we derive, as a byproduct, a nice mathematical identity
 \begin{gather*}
H_{n+m}(z) =   \sum_{j=0}^{\min(n,m)}   (-2)^j \frac {n! m!}{j! (n-j)! (m-j)!} H_{n-j}(z) H_{m-j}(z)  .
 \end{gather*}

In contrast to the standard PU oscillator where the Jordan blocks had inf\/inite dimension,  their
dimension is f\/inite here. We have a single vacuum state with energy $E = \Omega$, the Jordan block of dimension 2 at the
level $E = 2\Omega$, the Jordan block of dimension 3 at the level $E= 3\Omega$, etc.\ At each level, there is
a f\/inite number of dif\/ferent nonstationary solutions to the time-dependent Schr\"odinger equation and
 one cannot construct, as we did before,
bounded in $x,t$ combinations that have the meaning of continuous spectrum wave functions. As a result,
unitarity is violated\footnote{In~\cite{BM2}, a certain norm was constructed that is conserved during evolution. As a result,
Bender and Mannheim claimed that the system with the Hamiltonian (\ref{Hamrot}) in the equal frequency limit
is unitary. However, their norm  is nilpotent and, as was discussed at the beginning of Section~\ref{section3}, it amounts
to projecting the original Hilbert space  onto a complicated
unnaturally def\/ined subspace.  Even though the dynamics in the latter is unitary
(and the Hamiltonian loses its Jordan block structure
and is Hermitian), this is not the unitarity and Hermiticity in the standard meaning of these words.}.

As was mentioned, the real problem associated with the ghosts
 is the quantum and/or classical collapse, which might  appear only in interacting theory.
It is not so easy to include interactions and analyze their ef\/fects in the Bender--Mannheim approach.
First, it is not clear whether the Hamiltonian (\ref{Hamrot}) with the interaction term like $\lambda y^4$
is still crypto-Hermitian. The canonical transformations (\ref{canoncompl}) [or, which is equivalent, the similarity
transformation (\ref{simil})] kill the complexity $-ixp_y$, but
bring about new complexities coming from the interaction term $\propto y^4$.
These complexities have the form
$\Delta H_{\rm diag} \propto i \lambda (\Omega_1^2 P_1 X_2^3 - P_1^3 X_2)$.
We failed to f\/ind a modif\/ied canonical transformation that would kill all  complexities.

 Even if one succeeds in f\/inding a nonlinear crypto-Hermitian generalization of the Hamiltonian~(\ref{Hamrot}),
it would be dif\/f\/icult to analyze its spectrum and to f\/ind out whether the ground state is still present
there.
In particular, it would be dif\/f\/icult to implement variational estimates due to a complicated nonlocal norm.

\section{Conclusions}\label{section6}
Our main message is that the ghosts (the absence of the ground state in the spectrum), do not represent a serious problem
for {\it free} theory. Thus, it is not necessary to cope with them there. The spectrum of the free PU oscillator runs from
$-\infty$ to $+\infty$ representing a pure point spectrum that is dense everywhere when $\Omega_1 \neq \Omega_2$ and a
continuous
spectrum when $\Omega_1 = \Omega_2$. The latter can also be interpreted via inf\/inite-dimensional Jordan blocks
that appear in the
limit $\Omega_1 \to \Omega_2$~\cite{dno}. In spite of the absence of the ground state, unitarity is preserved.

Generically, interactions break unitarity due to quantum collapse  phenomenon. However, there are certain
 special cases when the theory
 involves neither classical nor quantum collapse. A particularly interesting example is the exactly soluble
nonlinear Hamiltonian
(\ref{HRobert}) that  appears in the context of supersymmetric higher derivative quantum mechanics \cite{Robert}\footnote{The Hamiltonian (\ref{HRobert}) had continuous spectrum, like the free PU oscillator with equal frequencies.
Another interesting nonlinear higher-derivative Hamiltonian,
\[
H = pP +D (\Omega^2 x + \lambda x^3) - \frac \gamma 2 (D^2 + P^2) ,
\]
has the pure point spectrum that is dense everywhere, like the PU oscillator at dif\/ferent frequencies.}.

By analytical continuation to complex coordinates, it is possible to consider a nonstandard realization of the
free PU Hamiltonian with the positive def\/inite spectrum (\ref{specrot12}). When $\Omega_1 \neq \Omega_2$, a unitary evolution
operator can be def\/ined. The presence of the ground state may be considered as a certain advantage of this realization
compared to the standard one. However,
 \begin{enumerate}\itemsep=0pt
\item It is misleading to say that choosing the nonstandard realization ``solves the ghost problem''. The analysis
has been performed only for  free theory and,
in the free case, there is no serious problem to solve anyway.
\item In contrast to the standard one, the nonstandard PU oscillator at equal frequencies is not unitary.
\item As far as the original problem with the Lagrangian (\ref{PUosc}) is concerned, the nonstandard realization
does not look natural: {\it (i)} it is not so natural to assume that the coordinate $q$ is imaginary while its time derivative
$x = \dot q$ is real;
{\it (ii)} the norm in the  space $(y \equiv -iq, \ x)$ that is preserved during evolution has a complicated nonlocal structure.
 \item It is not known yet what happens in the framework of nonstandard realization when interactions
are included.
 \end{enumerate}

\subsection*{Acknowledgements}

I acknowledge warm hospitality at AEI in Golm, where this work was f\/inished and thank
 P.~Mannheim for useful correspondence.

\pdfbookmark[1]{References}{ref}

\end{document}